\documentclass[10pt,reqno]{article}
\usepackage{amssymb, amsmath, amsthm, amsfonts, amscd}
\usepackage[english]{babel}
\usepackage{graphicx, epsfig}
\newtheorem{theorem}{Theorem}[section]

\newtheorem{prop}[theorem]{Proposition}

\newcommand{\nm}{\noalign{\smallskip}}

\newcommand{\Om}{\Omega}
\newcommand{\Bx}{\mathbf{x}}
\newcommand{\By}{\mathbf{y}}

\newcommand{\RR}{\mathbb{R}}

\newcommand{\NN}{\mathbb{N}}
\newcommand{\ZZ}{\mathbb{Z}}

\newcommand{\p}{\partial}
\newcommand{\Ga}{\alpha}
\newcommand{\Gb}{\beta}
\newcommand{\Gl}{\lambda}
\newcommand{\Gs}{\sigma}
\newcommand{\pd}[2]{\frac {\p #1}{\p #2}}
\newcommand{\ds}{\displaystyle}
\newcommand{\eqnref}[1]{(\ref {#1})}

\renewcommand{\qed}{\hfill $\Box$ \medskip}
\newcommand{\beq}{\begin{equation}}
\newcommand{\eeq}{\end{equation}}
\numberwithin{equation}{section}
\numberwithin{figure}{section}

\begin{document}

\title{Enhancement of Near Cloaking Using Generalized Polarization Tensors Vanishing Structures. Part I: The Conductivity
Problem\thanks{\footnotesize
This work was supported  by National Institute for Mathematical
Sciences (2010 Thematic Program, TP1003), ERC Advanced Grant
Project MULTIMOD--267184, Korea Research Foundation through grant
KRF-2008-220-C00002, and NRF grants No. 2009-0090250,
2010-0017532, and 2010-0004091, and grants from Inha University.}}

\author{Habib Ammari\thanks{\footnotesize Department of Mathematics and Applications, Ecole Normale Sup\'erieure,
45 Rue d'Ulm, 75005 Paris, France (habib.ammari@ens.fr).}  \and
Hyeonbae Kang\thanks{Department of Mathematics, Inha University,
Incheon 402-751, Korea (hbkang@inha.ac.kr, hdlee@inha.ac.kr).}
\and Hyundae Lee\footnotemark[3] \and Mikyoung
Lim\thanks{\footnotesize Department of Mathematical Sciences,
Korean Advanced Institute of Science and Technology, Daejeon
305-701, Korea (mklim@kaist.ac.kr). }}

\maketitle

\begin{abstract}
The aim of this paper is to provide an original method of
constructing very effective near-cloaking structures for the
conductivity problem. These new structures are such that their
first Generalized Polarization Tensors vanish. We show that this in particular
significantly enhances the cloaking effect. We then present some numerical examples of
Generalized Polarization Tensors vanishing structures.
\end{abstract}

\noindent {\footnotesize {\bf AMS subject classifications.} 35R30,
35B30}

\noindent {\footnotesize {\bf Key words.} cloaking, generalized
polarization tensor, inverse conductivity problem}

\section{Introduction}
The central problem in the electrical impedance tomography is to
reconstruct the unknown conductivity distribution of a conductor
using the boundary measurements, the Dirichlet-to-Neumann (DtN)
map. Even if unique identifiability by the DtN map holds for wide
class of conductivity distribution (\cite{AP, Na95, SU} to cite
only a few), Greenleaf {\it et al.} \cite{glu} found a (singular)
conductivity distribution whose DtN map is exactly the same as the
one associated to the constant conductivity distribution. They use
a change of variables scheme to create the desired  conductivity
distribution. They push forward the material constant by the
transformation blowing up a point to a ball or a disk. It turned
out that the change of variables scheme can be applied to
cloaking: Pendry {\it et al.} \cite{pendry} and Leonhardt \cite{leonhardt} used similar idea to initiate the research on cloaking.
Cloaking is to make a target invisible with respect to  probing by
electromagnetic waves. Since then, extensive
  work has been produced on cloaking in the context of conductivity and electromagnetism. We refer to
  \cite{GKLU} (also \cite{BL}) for recent development on the cloaking. It is worth mentioning that there
  is yet another kind of cloaking in which  the cloaking region is outside the cloaking device, for instance,
  anomalous localized resonance \cite{MN_PRSA_06, MNMP_PRSA_05}.

The change of variables based cloaking method uses the singular
transformation to boost the material property so that it makes a
cloaking region look like a point to outside measurements.
However, this transformation induces the singularity of material
constants in the transversal direction (also in the tangential
direction in two dimensions), which invokes the difficulty both in
the theory and applications. To overcome this weakness, so called
`near cloaking' is naturally considered, which is the regularization or the approximation of singular cloaking. In \cite{kohn1}, instead of the singular transformation, Kohn {\it et al.} use a regular one to
push forward the material constant in the conductivity equation, in which a
small ball (of radius $\rho)$ is blew up to the cloaking region. They estimate that this
near-cloaking can be approximated to the perfect one with the order of $\rho^d$ in the space of dimension $d$.

The purpose of this paper is to propose a new cancelation
technique in order to achieve enhanced near-invisibility. Our approach is
based on the multi-coating which cancels the generalized
polarization tensors (GPTs) of the cloaking device. We first
design a structure coated around an inclusion to have vanishing
GPTs of lower orders and show that the order of perturbation due
to a small inclusion can be reduced significantly. We then obtain
near-cloaking structure by pushing forward the multi-coated
structure around a small object via the usual blow-up
transformation.  For the conductivity equation, we show that the
order of near-cloaking is $\rho^{2N}$ using $N$ coatings, which is
a significant improvement over $\rho^2$ approximation obtained in
\cite{kohn1}. We give numerical examples for the material
parameters and the thickness of the layers for the GPT vanishing structures.

This paper is organized as follows. In the next section we derive the multi-polar expansion of the solution to the conductivity equation which is slightly different from the usual one, and define the (contracted) generalized polarization tensors. In section 3 we characterize the GPT vanishing structures. In section 4 we show that the near-cloaking is enhanced (to $\rho^{2N}$) if the GPT vanishing structure is used. In section 5 we present some numerical examples of the GPT vanishing structures. We end this paper with a brief conclusion.

Even though we consider only two dimensional conductivity equation in this paper, the same argument can be applied to the equation in three dimensions. The multi-coating technique developed in this paper can be applied to the Helmholtz equation to enhance the near cloaking obtained in \cite{nguyen, kohn2, liu}. The results for the Helmholtz equation will be presented in the forthcoming paper.

%%%%%%%%%%%%%%%%%%%%%%%%%%%%%%%%%%%%%%%%%%%%%%%%%%%%%%%%%%%%%%
\section{Far-field behavior of the solution}
%%%%%%%%%%%%%%%%%%%%%%%%%%%%%%%%%%%%%%%%%%%%%%%%%%%%%%%%%%%%%%

Let $\Om$ be a domain in $\RR^2$ containing $0$ possibly with multiple components with Lipschitz boundary.
For a given harmonic function $H$ in $\RR^2$, consider
 \beq\label{CP1}
 \ \left \{
\begin{array}{l}
\ds \nabla  \cdot \Bigr( \sigma_0 \chi(\RR^2 \setminus \overline{\Om}) + \sigma
\chi(\Om) \Bigr) \nabla u =0 \quad \mbox{in } \RR^2,
\\ \nm \ds u(\Bx)- H(\Bx) = O(|\Bx|^{-1}) \quad\mbox{as } |\Bx|
\rightarrow \infty,
\end{array}
\right .
 \eeq
where $\sigma_0$ and $\sigma$ are conductivities (positive
constants) of $\RR^2 \setminus \Om$ and $\Om$, respectively. The
solution $u$ to \eqnref{CP1} admits the multipolar expansion
\cite{book2}
 \beq\label{CP2}
 (u-H)(\Bx) = \sum_{|\Ga|,|\Gb|=1}^\infty\frac{(-1)^{|\Ga|}}{\Ga!\Gb!}\p^\Ga_\Bx\Gamma(\Bx)M_{\Ga\Gb}\p^\Gb H(0), \quad |\Bx| \to \infty,
 \eeq
where $M_{\Ga\Gb}= M_{\Ga\Gb}(\Om, \frac{\Gs}{\Gs_0})$ are the
Generalized Polarization Tensors (GPTs) associated with the
inclusion $\Om$ and the conductivity contrast $\frac{\Gs}{\Gs_0}$
and $\Gamma(\Bx)$ is the fundamental solution of the Laplacian,
{\it i.e.},
 $$
 \Gamma(\Bx)= \frac{1}{2\pi} \ln |\Bx|.
 $$
Here and throughout this paper $\Ga=(\Ga_1, \Ga_2)$ and
$\Gb=(\Gb_1, \Gb_2)$ are multi-indices and $|\Ga|=\Ga_1+\Ga_2$.

We seek an expression of the multipolar expansion which is
slightly different from \eqnref{CP2}. For multi-indices $\alpha$,
$\beta$ with ${|\Ga|=n}$, ${|\Gb|=n}$, define $(a^n_{\Ga})$ and
$(b^n_{\Gb})$ by
 \beq\label{harcoe}
 \sum_{|\Ga|=n} a^n_{\Ga} \Bx^{\Ga}= r^n\cos n\theta \quad\mbox{and}\quad
 \sum_{|\Gb|=n} b^n_{\Gb} \Bx^{\Gb}= r^n \sin n\theta,
 \eeq
and define
 \begin{align}
 \ds M_{mn}^{cc}= \sum_{|\Ga|=m} \sum_{|\Gb|=n} a^m_{\Ga} a^n_{\Gb} M_{\Ga\Gb}, \\
 \ds M_{mn}^{cs}= \sum_{|\Ga|=m} \sum_{|\Gb|=n} a^m_{\Ga} b^n_{\Gb} M_{\Ga\Gb},\\
 \ds M_{mn}^{sc}= \sum_{|\Ga|=m} \sum_{|\Gb|=n} b^m_{\Ga} a^n_{\Gb} M_{\Ga\Gb}, \\
 \ds M_{mn}^{ss}= \sum_{|\Ga|=m} \sum_{|\Gb|=n} b^m_{\Ga} b^n_{\Gb} M_{\Ga\Gb}.
 \end{align}
We call these coefficients the contracted GPTs.

Using the expansion of $\ln |\Bx-\By|$, we have
\beq
 \sum_{|\Ga|=n}\frac{(-1)^{|\Ga|}}{\Ga!}\p_\Bx^\Ga\Gamma(\Bx)\By^\Ga
 =\frac{-1}{2\pi n}\left[\frac{\cos n\theta_\Bx}{r_\Bx^n}r_\By^n\cos n\theta_\By
 +\frac{\sin n\theta_\Bx}{r_\Bx^n}r_\By^n\sin n\theta_\By\right],\quad n>0,
 \eeq
where $\Bx=r_\Bx(\cos\theta_\Bx,\sin\theta_\Bx)$ and $\By=r_\By(\cos\theta_\By,\sin\theta_\By)$, which is valid if $|\Bx| \gg 1$ and $\By \in \p\Om$,
 and hence
\beq \label{eq234}
\frac{(-1)^{|\Ga|}}{\Ga!}\p_\Bx^\Ga\Gamma(\Bx)=\frac{-1}{2\pi n}\left[ a^n_{\Ga}\frac{\cos n\theta_\Bx}{r_\Bx^n}+ b^n_{\Ga}\frac{\sin n\theta_\Bx}{r_\Bx^n}\right].
\eeq

If the harmonic function $H$ admits the expansion
 $$
 H(\Bx) = H(0)+\sum_{n=1}^\infty r^n \bigr(a_n^c(H)\cos n\theta + a_n^s(H)\sin n\theta)\bigr)
 $$
with $\Bx=(r\cos\theta, r\sin\theta)$, then we have
\beq \label{eq235}
\frac{\p^\Ga H(0)}{\Ga!}= a_n^c(H)a^n_\Ga + a_n^s(H)b^n_\Ga.
\eeq

Plugging \eqnref{eq234} and \eqnref{eq235} into \eqnref{CP2} we have
 the following formula
\begin{align}
(u-H)(\Bx) & = -\sum_{m=1}^\infty\frac{\cos m\theta}{2\pi mr^m}\sum_{n=1}^\infty
\bigr(M_{mn}^{cc}a_n^c(H)+M_{mn}^{cs}a_n^s(H)\bigr) \nonumber \\
& \quad -\sum_{m=1}^\infty\frac{\sin m\theta}{2\pi m
r^m}\sum_{n=1}^\infty
\bigr(M_{mn}^{sc}a_n^c(H)+M_{mn}^{ss}a_n^s(H)\bigr) \quad\mbox{as
} |\Bx| \to \infty. \label{expan2}
\end{align}

We emphasize that \eqnref{expan2} is valid even if $\Gs$ (the conductivity of $\Om$) is not
a constant but a variable. In the next section we will design a conductivity distribution $\Gs$ of $\Om$ so that $M_{mn}^{cc}=M_{mn}^{cs}=M_{mn}^{sc}=M_{mn}^{ss}=0$ for all $m, n \le N$ for a given integer $N$. We call such a conductivity distribution {\it GPT-vanishing structure} or {\it coating} of order $N$.

%%%%%%%%%%%%%%%%%%%%%%%%%%%%%%%%%%%%%%%%%%%%%%%%%%%%%%%%%%%
\section{GPT vanishing structures}\label{multicoatcond}
%%%%%%%%%%%%%%%%%%%%%%%%%%%%%%%%%%%%%%%%%%%%%%%%%%%%%%%%%%%

To obtain GPT-vanishing structures, we use a disc with multiple
coatings. The idea comes from Hashin's neutral inclusion which is
a disc with a single coating \cite{milton_book}. The special
property of the neutral inclusion is that it does not perturb the
uniform fields outside the inclusion, which is equivalent to the first order
polarization tensors of the inclusion vanishing. In other words, Hashine's neutral
inclusion is a GPT-vanishing structure of order 1.

Let $\Om$ be a disk of radius $r_1$. For a positive integer $N$, let $0< r_{N+1} < r_{N} < \ldots < r_1$ and define
 \beq\label{Aj}
 A_j := \{ r_{j+1} < r \le r_{j} \}, \quad j=1,2,\ldots, N.
 \eeq
Let $A_0=\RR^2 \setminus \Om$ and $A_{N+1}=\{ r \le r_{N+1} \}$.
Set $\sigma_j$ to be the conductivity of $A_j$ for $j=1,2,\ldots, N+1$, and $\sigma_0=1$. Let
 \beq\label{condis}
 \sigma = \sum_{j=0}^{N+1} \sigma_j \chi(A_j) .
 \eeq
Let $M_{mn}^{cc}[\sigma]$, etc, denote the (contracted) GPTs associated with $\Gs$ (and $\Om$).
Because of the symmetry of the disc, one can easily see that
 \beq
 M_{mn}^{cs}[\Gs]= M_{mn}^{sc}[\Gs] =0\quad\mbox{for all }m,n,
 \eeq
 \beq
 M_{mn}^{cc}[\Gs]=M_{mn}^{ss}[\Gs]=0 \quad\mbox{if } m \neq n,
 \eeq
and
 \beq
 M_{nn}^{cc}[\Gs]=M_{nn}^{ss}[\Gs] \quad\mbox{for all } n.
 \eeq
Let $M_n=M_{nn}^{cc}$, $n=1,2,\ldots$, for the simplicity of notation.

To compute $M_{k}$, we look for solutions $u_k$ to
 \beq \label{eq3456}
 \nabla \cdot \sigma \nabla u=0 \quad \mbox{in }\RR^2
 \eeq
of the form
 \beq
 u_k(\Bx) = a_j^{(k)} r^k \cos k\theta + \frac{b_j^{(k)}}{r^k} \cos k\theta \quad \mbox{in } A_j, \ \ j=0, 1, \ldots,  N+1,
 \eeq
with $a_0^{(k)}=1$ and $b_{N+1}^{(k)}=0$. Then $u_k$ is the solution to \eqnref{CP1} with $H(\Bx) = r^k\cos k\theta$, and satisfies
 \beq
 (u-H)(\Bx) = \frac{b_0^{(k)}}{r^k} \cos k\theta \quad\mbox{as } |\Bx| \to \infty.
 \eeq
Hence, we have
 \beq
 M_{k} =-2\pi k b_0^{(k)}.
 \eeq

We observe that the solution to \eqnref{eq3456} is given by
 \beq
 u_k(re^{i\theta}) = \left(r^k - \frac{M_k}{2\pi k r^k}\right) \cos k\theta \quad \mbox{in } A_0.
 \eeq
Thus $$r^k \pm \frac{M_k}{2\pi k r^k} \neq 0 \quad \mbox{ for any
} r \ge 2,$$ since otherwise $u_k(\Bx)=0$ or
$u_k(\Bx)=\mbox{constant}$ in $|\Bx| \le r$ for some $r \ge 2$.
Thus we have
 $$
 r^k - \frac{M_k}{2\pi k r^k} >0 \quad\mbox{and}\quad r^k + \frac{M_k}{2\pi k r^k}>0 \quad\mbox{for any } r \ge 2,
 $$
and hence
\begin{align} \label{boundM}
|M_k| \le 2\pi k 2^{2k} \quad\mbox{for all } k.
\end{align}

The transmission conditions on the interface $\{ r=r_j \}$ are given by
 \begin{align}
 a_j^{(k)} r_j^k + \frac{b_j^{(k)}}{r_j^k} &= a_{j-1}^{(k)} r_{j}^k + \frac{b_{j-1}^{(k)}}{r_{j}^k} , \\
 \sigma_j \left( a_j^{(k)} r_j^{k-1} - \frac{b_j^{(k)}}{r_j^{k+1}} \right) &= \sigma_{j-1} \left( a_{j-1}^{(k)} r_{j}^{k-1} - \frac{b_{j-1}^{(k)}}{r_{j}^{k+1}} \right).
 \end{align}
Thus we have
 \beq\label{Pmatrix}
 \begin{bmatrix}  a_j^{(k)} \\ b_j^{(k)} \end{bmatrix}
 = \frac{1}{2 \sigma_j} \begin{bmatrix}  \sigma_j + \sigma_{j-1} & (\sigma_j - \sigma_{j-1}) r_j^{-2k} \\ (\sigma_j - \sigma_{j-1}) r_j^{2k} & \sigma_j + \sigma_{j-1} \end{bmatrix} \begin{bmatrix}  a_{j-1}^{(k)} \\ b_{j-1}^{(k)} \end{bmatrix} ,
 \eeq
and hence
 \beq
 \begin{bmatrix}  a_{N+1}^{(k)} \\ 0 \end{bmatrix}
 = \prod_{j=1}^{N+1} \frac{1}{2 \sigma_j} \begin{bmatrix}  \sigma_j + \sigma_{j-1} & (\sigma_j - \sigma_{j-1}) r_j^{-2k} \\ (\sigma_j - \sigma_{j-1}) r_j^{2k} & \sigma_j + \sigma_{j-1} \end{bmatrix} \begin{bmatrix} 1 \\ b_0^{(k)} \end{bmatrix} .
 \eeq
Let
 \beq\label{Pmatrix_total}
 P^{(k)}=\begin{bmatrix}p_{11}^{(k)} & p_{12}^{(k)}\\
  p_{21}^{(k)} & p_{22}^{(k)}
  \end{bmatrix}
  :=\prod_{j=1}^{N+1} \frac{1}{2 \sigma_j} \begin{bmatrix}  \sigma_j + \sigma_{j-1} & (\sigma_j - \sigma_{j-1}) r_j^{-2k} \\ (\sigma_j - \sigma_{j-1}) r_j^{2k} & \sigma_j + \sigma_{j-1} \end{bmatrix}.
 \eeq
Then, we have
 \begin{equation}\label{Mk1}
b_0^{(k)}= -\frac{p^{(k)}_{21}}{p^{(k)}_{22}}.
 \end{equation}

Note that $M_{k}=0$ if and only if $P^{(k)}$ is an upper triangular matrix, {\it i.e.}, $p^{(k)}_{21}=0$. Let
 \beq
 \Gl_j := \frac{\sigma_j - \sigma_{j-1}}{\sigma_j + \sigma_{j-1}}, \quad j=1, \ldots,
 N. \label{const3}
 \eeq

We now characterize GPT-vanishing structures.
\begin{prop}
If there are non-zero constants $\lambda_1, \ldots, \lambda_{N+1}$
($|\lambda_j| < 1$) and $r_1 > \ldots > r_{N+1} >0$ such that
\beq\label{fincloak} \prod_{j=1}^{N+1}
\begin{bmatrix}
\ds 1 & \ds \lambda_j r_j^{-2k} \\
\ds \lambda_j r_j^{2k} & 1
\end{bmatrix}
\mbox{ is an upper triangular matrix for }k=1, 2, \ldots, N, \eeq
then $(\Om, \Gs)$, given by \eqnref{Aj}, \eqnref{condis}, and
\eqnref{const3},  is a GPT-vanishing structure with $M_{k}=0$ for
$k \le N$. More generally, if that there are non-zero constants
$\lambda_1, \Gl_2, \Gl_3, \ldots$ ($|\lambda_j| < 1$) and $r_1 >
r_2 > r_3 > \ldots$ such that $r_n$ converges to a positive
number, say $r_\infty >0$, and \beq\label{infincloak}
\prod_{j=1}^{\infty}
\begin{bmatrix}
\ds 1 & \ds \lambda_j r_j^{-2k} \\
\ds \lambda_j r_j^{2k} & 1
\end{bmatrix}
\mbox{ is an upper triangular matrix for every } k, \eeq then
$(\Om, \Gs)$, given by \eqnref{Aj}, \eqnref{condis}, and
\eqnref{const3},  is a GPT-vanishing structure with $M_{k}=0$ for
all $k$. \end{prop}

Note that \eqnref{fincloak} and \eqnref{infincloak} are nonlinear
equations. For example, if $N=3$, equation \eqnref{fincloak} is
simple and reduces to
 \beq
 \lambda_1r_1^{2k}+\lambda_2r_2^{2k}+\lambda_3r_3^{2k}+\lambda_1\lambda_2\lambda_3 r_1^{2k}r_2^{-2k}r_3^{2k}=0,\quad k=1,2,3.
 \eeq
It is quite easy to show that it admits infinitely many solutions
$\lambda_1, \Gl_2, \Gl_3$ ($|\lambda_j| < 1$) and $r_1 > r_2 >
r_{3} >0$. However, as $N$ gets larger, solving analytically
equation \eqnref{fincloak} seems too complicated, and even proving
existence of solutions to \eqnref{fincloak} or \eqnref{infincloak}
seems to be quite challenging. We present a simple numerical
method to find the GPT-vanishing structures in Section
\ref{sectnum}, which is also important from a practical point of
view. These numerical evidences show us that \eqnref{fincloak} has
solutions, even though we are not able to prove it.

%%%%%%%%%%%%%%%%%%%%%%%%%%%%%%%%%%%%%%%%%%%%%%%%%%%%%
\section{Near-cloaking using GPT-vanishing structures}
%%%%%%%%%%%%%%%%%%%%%%%%%%%%%%%%%%%%%%%%%%%%%%%%%%%%%

In this section we achieve enhanced near-cloaking by using GPT-vanishing
structures.

Let $B_r$ be the disk centered at the origin with radius $r$ and
$B=B_1$. Let $A_j$, $j=0, 1, \ldots, N+1$ be defined by
\eqnref{Aj} with $r_1=2$ and $r_{N+1}\ge 1$. Let $\Gs$ be the
conductivity distribution defined by \eqnref{condis}. For a given
domain $\Omega$, we denote the DtN map of $\Omega$ with the
conductivity $\sigma$ as $\Lambda_{\Omega}[\sigma]$, which is
given by
 \beq
 \Lambda_{\Om} [\sigma](f) = \Gs \pd{u}{\nu} \bigg|_{\p\Om}
 \eeq
where $u$ is the solution to
 \begin{equation}
 \begin{cases}
 \ds\nabla\cdot \Gs \nabla u  =0 \quad & \mbox{in } \Om, \\
  \nm \ds u=f \quad & \mbox{on } \p\Om.
 \end{cases}
 \end{equation}

\begin{prop} \label{mainprop}
Let $\sigma$ be a conductivity profile on $\RR^2$ defined as \eqnref{condis} and for a small positive constant let
 \beq
 \Psi_{\frac{1}{\rho}}(\Bx)=\frac{1}{\rho}\Bx, \quad \Bx\in\RR^2.
 \eeq
Then the following holds for $k=0, 1, 2, \ldots$ and $s>0$
 \beq\label{firstcomp}
 \Bigr(\Lambda_{B_s}[\sigma\circ\Psi_{\frac{1}{\rho}}]-\Lambda_{B_s}[1]\Bigr)(e^{\pm ik\theta})
=\frac{2ks^{-1}\rho^{2k}M_k[\sigma]}{2\pi k s^{2k}-M_k[\sigma] \rho^{2k}}e^{\pm ik\theta} .
 \eeq
\end{prop}

Before proving Proposition \ref{mainprop}, let us make a few remarks.
If $\Gs$ is a GPT-vanishing structure of order $N$, {\it i.e.}, $M_k=0$ for all $k \le N$, then
\eqnref{firstcomp} shows that
 \beq\label{lowoscil}
 \Bigr(\Lambda_{B_s}[\sigma\circ\Psi_{\frac{1}{\rho}}]-\Lambda_{B_s}[1]\Bigr)(e^{\pm ik\theta})
 =0 \quad\mbox{for all } |k| \le N.
 \eeq
In other words, $\Lambda_B[\sigma\circ\Psi_{\frac{1}{\rho}}]$ and $\Lambda_B[1]$ cannot be distinguished by slowly oscillating Dirichlet data. Moreover, the complete GPT-vanishing structure \eqnref{infincloak} is achieved, then
 \beq\label{lowoscil2}
 \Bigr(\Lambda_{B_s}[\sigma\circ\Psi_{\frac{1}{\rho}}]-\Lambda_{B_s}[1]\Bigr)(e^{\pm ik\theta})
 =0 \quad\mbox{for all } k,
 \eeq
which would yield the perfect cloaking.

\medskip

\noindent{\it Proof of Proposition \ref{mainprop}}.
 If $u$ is the solution to
 \begin{equation}
 \begin{cases}
 \ds\nabla\cdot(\sigma\circ\Psi_{\frac{1}{\rho}})\nabla u =0 \quad & \mbox{in } B_{s} , \\
  \nm \ds u=f \quad & \mbox{on } |\Bx|=s,
 \end{cases}
 \end{equation}
then $\tilde{u}:=u\circ\Psi_\rho$ satisfies
 \begin{equation}\label{tildeu}
 \begin{cases}
 \ds\nabla\cdot\sigma\nabla \tilde u  =0 \quad & \mbox{in } B_{\frac{s}{\rho}} , \\
 \nm \ds \tilde u=f\circ\Psi_\rho \quad & \mbox{on } |\By|=\frac{s}{\rho}.
 \end{cases}
 \end{equation}
 Moreover, we have
$$\pd{\tilde{u}}{\nu}\Bigr|_{|\By|=\frac{s}{\rho}}(\By) = \rho\pd{u}{\nu}\Bigr|_{|\Bx|=s}(\rho \By).$$ Therefore,
\begin{equation}\label{rhoscale}
\Lambda_{B_{\frac{s}{\rho}}}[ \sigma](f\circ\Psi_\rho)
= \rho \Lambda_{B_s}[\sigma\circ\Psi_\frac{1}{\rho}](f) \circ \Psi_\rho   .
\end{equation}

To compute $\Lambda_{B_{\frac{s}{\rho}}}[ \sigma]$, set $f(e^{i\theta})= e^{ik\theta}$ for a fixed $k\in\NN$ ($k \ge 0$).
Then, the solution $\tilde u_k$ to \eqnref{tildeu} is given by
 $$
 \tilde{u}_k(\Bx) = a_j^{(k)} r^k e^{ik\theta} + \frac{b_j^{(k)}}{r^k} e^{ik\theta} \quad \mbox{in } A_j, \ \ j=0, 1, \ldots, N+1,
 $$
with \begin{equation}\label{a0b0} a_{0}^{(k)}s^k\rho^{-k}+
b_{0}^{(k)}s^{-k}\rho^{k} =1\end{equation} and $b_{N+1}^{(k)}=0$.
From \eqnref{Pmatrix} and \eqnref{Pmatrix_total}, we get
 \beq\nonumber
 \begin{bmatrix}  a_{N+1}^{(k)} \\ 0 \end{bmatrix}
 = \begin{bmatrix} p_{11}^{(k)} & p_{12}^{(k)} \\
 p_{21}^{(k)} & p_{22}^{(k)} \end{bmatrix} \begin{bmatrix} a_0^{(k)} \\ b_0^{(k)} \end{bmatrix},
 \eeq
and hence $$b_0^{(k)} -\frac{p^{(k)}_{21}}{p^{(k)}_{22}}a_0^{(k)} =-
\frac{M_k[\sigma]}{2\pi k} a_0^{(k)}.$$ Substituting it into
\eqnref{a0b0}, we have
$$b_0^{(k)} = \frac{-M_k[\sigma]\rho^k s^k}{2\pi k s^{2k}-M_k[\sigma] \rho^{2k}}.$$
and we obtain
\begin{align*}
\Lambda_{B_{\frac{s}{\rho}}}[ \sigma](f\circ\Psi_\rho)&=\pd{\tilde u_k}{\nu}\Bigr|_{|\By|=\frac{s}{\rho}}\\
&=k\frac{\rho}{s}(a_0^{(k)}s^k\rho^{-k} -
b_0^{(k)}s^{-k}\rho^{k})e^{ik\theta}\\
&=k\frac{\rho}{s}(a_0^{(k)}s^k\rho^{-k} +
b_0^{(k)}s^{-k}\rho^{k})e^{ik\theta}-2k b_0^{(k)}\frac{\rho^{k+1}}{s^{k+1}}e^{ik\theta}\\
&=k\frac{\rho}{s} e^{ik\theta} -2k b_0^{(k)}\frac{\rho^{k+1}}{s^{k+1}}e^{ik\theta}.
\end{align*}
From \eqnref{rhoscale} we have
\begin{align*}
\Lambda_{B_s}[\sigma\circ\Psi_\frac{1}{\rho}](f)&=\frac{k}{s}e^{ik\theta} - 2k b_0^{(k)}\frac{\rho^{k}}{s^{k+1}}e^{ik\theta}\\
&=\Lambda_{B_s}[1](e^{ik\theta}) +\frac{2ks^{-1}\rho^{2k}M_k[\sigma] }{2\pi k s^{2k}-M_k[\sigma] \rho^{2k}}e^{ik\theta}.
\end{align*}
Thus we get \eqnref{firstcomp}. The same argument works for $k<0$ as well. The proof is complete.
\qed

If $f$ admits Fourier expansion $f(e^{i\theta})=\sum_{k\in\ZZ}f_ke^{ik\theta}$, then we have from \eqnref{firstcomp}
 \beq
 \Bigr(\Lambda_B[\sigma\circ\Psi_{\frac{1}{\rho}}]-\Lambda_B[1]\Bigr)(f)
 =\sum_{k\in\ZZ}\frac{2|k|\rho^{2|k|}s^{-1}M_{|k|}[\sigma]}{2\pi |k|s^{2|k|}-\rho^{2|k|}M_{|k|}[\sigma] }f_ke^{ik\theta}.
 \eeq
For the GPT-vanishing structure of order $N$, we have $M_k =0$ for all $k \le N$ and from \eqnref{boundM} that
 $$
 \rho^{2k}|M_k| \leq 2\pi k (2\rho)^{2k} \quad \mbox{for all }k >N.
 $$
Since $\rho \ll 1$, we obtain
 \beq\label{approxvanish}
 \rho^{2k}|M_k|\leq C \rho^{2N+2} \quad \mbox{for all }k\in \NN
 \eeq
for some constant $C$ independent of $k$, and hence
 \beq
 \Bigr\|\Lambda_{B_s}[\sigma\circ\Psi_\frac{1}{\rho}]-\Lambda_{B_s}[1]\Bigr\|\leq C \rho^{2N+2},
 \eeq
where the norm is the operator norm of $H^{1/2}(\p B_s)$ into
$H^{-1/2}(\p B_s)$. Here $H^{1/2}(\p B_s)$ is the usual Sobolev
space of order $1/2$ on $\p B_s$ and $H^{-1/2}(\p B_s)$ is its
dual.

\medskip

We mention that the solutions for GPT-vanishing structure exist
numerically even when $\sigma_{N+1}= 0$, which is equivalent to
prescribe the insulating condition on the boundary of the inner
core. Some examples with $\sigma_{N+1}=0$ and $r_{N+1}=1$ are
given in the following section. In such structures, the
conductivity $\sigma\circ\Psi_\frac{1}{\rho}$ is $0$ in $|\Bx| \le
\rho$, fluctuates in $\rho < |\Bx| < 2\rho$, and is 1 in $|\Bx| >
2\rho$. In order to have near-cloaking device in $1<|\Bx|<2$ with
the core part $|\Bx| \le 1$ insulated as was considered in
\cite{kohn1}, one may use transformation as was done in
\cite{GKLU, pendry}.

For a given domain $\Omega$ and a subdomain $B \subset \Om$, we denote the DtN map of $\Omega$ with the conductivity
  $\sigma$ as $\Lambda_{\Omega,B}[\sigma]$, which is given by
 \beq
 \Lambda_{\Om,B} [\sigma](f) = \Gs \pd{u}{\nu} \bigg|_{\p\Om}
 \eeq
where $u$ is the solution to
 \begin{equation}
 \begin{cases}
 \ds\nabla\cdot \Gs \nabla u =0 \quad & \mbox{in } \Om \setminus \overline{B} , \\
 \ds \pd{u}{\nu}=0\quad & \mbox{on } \p B,\\
  \nm \ds u=f \quad & \mbox{on } \p\Om.
 \end{cases}
 \end{equation}

By \eqnref{firstcomp}, we have
 $$\Bigr(\Lambda_{B_2,B_\rho}[\sigma\circ\Psi_{\frac{1}{\rho}}]-\Lambda_{B_2,\emptyset}[1]\Bigr)(f)
=\sum_{k\in\ZZ}\frac{|k|\left(\frac{\rho}{2}\right)^{2|k|}M_{|k|}}{2\pi|k|-\left(\frac{\rho}{2}\right)^{2|k|}M_{|k|} }f_ke^{ik\theta}, $$
and hence by \eqnref{approxvanish}
 \beq\label{Btwocomp}
 \Bigr\|\Lambda_{B_2,B_\rho}[\sigma\circ\Psi_\frac{1}{\rho}]-\Lambda_{B_2,\emptyset}[1]\Bigr\|\leq C \rho^{2N+2}.
 \eeq
We define the transformation $F_{\rho}:B_2 \rightarrow B_2$ by
\begin{equation}\label{frho}
F_{\rho}(\Bx):= \begin{cases}
\ds \Bx \quad&\mbox{for }\frac{3}{2}\leq|\Bx|\leq2,\\
\ds\Bigr(\frac{3-3\rho}{3-2\rho}+\frac{1}{3-2\rho}|\Bx|\Bigr)\frac{\Bx}{|\Bx|} \quad&\mbox{for }\rho\leq|\Bx|\leq\frac{3}{2},\\
\ds\frac{\Bx}{\rho} &\mbox{for }|\Bx|\leq \rho.
\end{cases}
\end{equation}
Then one can easily see that
\begin{equation}
\Lambda_{B_2,B_\rho}[\sigma\circ\Psi_{\frac{1}{\rho}}]=\Lambda_{B_2,B_1}\left[(F_\rho)_*(\sigma\circ\Psi_{\frac{1}{\rho}})\right],
\end{equation}
where
 $$
 (F_{\rho})_*(\sigma\circ\Psi_{\frac{1}{\rho}}) =\frac{(DF_{\rho})(\sigma\circ\Psi_{\frac{1}{\rho}})(DF_{\rho})^T}{|\mbox{det}(DF_{\rho})|}\circ
 F_{\rho}^{-1}.
 $$
We then obtain the following theorem, which is the main result of this paper, from \eqnref{Btwocomp}.
\begin{theorem}
There exists a constant $C$ independent of $\rho$
such that
 \beq\label{Btwocomp2}
 \Bigr\|\Lambda_{B_2,B_1}\left[(F_\rho)_*(\sigma\circ\Psi_{\frac{1}{\rho}})\right]-\Lambda_{B_2,\emptyset}[1]\Bigr\|\leq C \rho^{2N+2} .
 \eeq
\end{theorem}

%%%%%%%%%%%%%%%%%%%%%%%%%%%%%%%%%%%%%%%
\section{Numerical examples} \label{sectnum}
%%%%%%%%%%%%%%%%%%%%%%%%%%%%%%%%%%%%%%%

In this section we present some numerical examples of $\Gs_1,
\ldots, \Gs_{N+1}$ and $r_1 > \ldots >
 r_{N+1}$ satisfying \eqnref{fincloak}.

We fix $N$ and $r_j = 2-\frac{j-1}{N}$ for $j=1,\dots,N+1$. We then solve the
following equation for ${\boldsymbol\sigma}=(\sigma_1,\dots,\sigma_{N+1})$
 \beq\label{Msigma}
 {\boldsymbol\sigma} \mapsto M_{k}[{\boldsymbol\sigma}]=0 \quad\mbox{for } k=1, \ldots, N.
 \eeq

Since \eqnref{Msigma} is a nonlinear equation, we solve it iteratively. Initially, $\sigma^{(0)}_j$ is set to be $2^{(-1)^j}$, $j=1,\dots,N+1$.
We iteratively modify ${\boldsymbol\sigma}^{(i)}=(\sigma^{(i)}_1,\dots,\sigma^{(i)}_{N+1})$ as
$$\boldsymbol\sigma^{(i+1)} = \boldsymbol\sigma^{(i)} - A_i^\dag \textbf{b}^{(i)},$$
where $A_i^\dag$ is the pseudoinverse of
$$
A_i:=\frac{\partial (M_{1},\dots,M_{N})}{\partial \boldsymbol\sigma}\Bigr|_{\boldsymbol\sigma =  \boldsymbol\sigma^{(i)}},$$
and
$$\textbf{b}^{(i)} = \left.\begin{bmatrix}
 M_{1} \\\vdots\\M_{N} \end{bmatrix}\right|_{\boldsymbol\sigma =  \boldsymbol\sigma^{(i)}}.$$

\begin{figure}[h!]
\begin{center}
\epsfig{figure=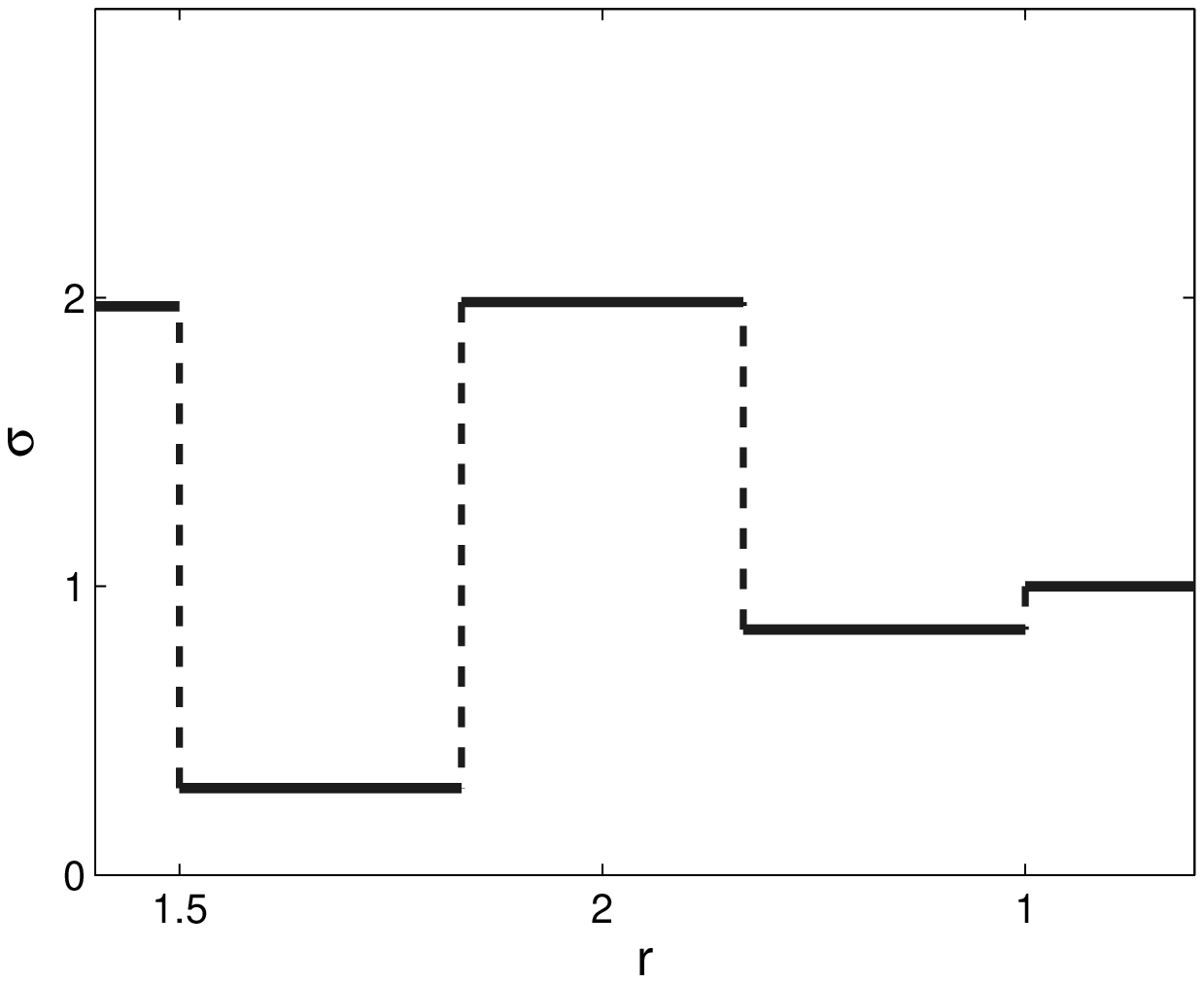, height=4cm}\hskip 1cm
\epsfig{figure=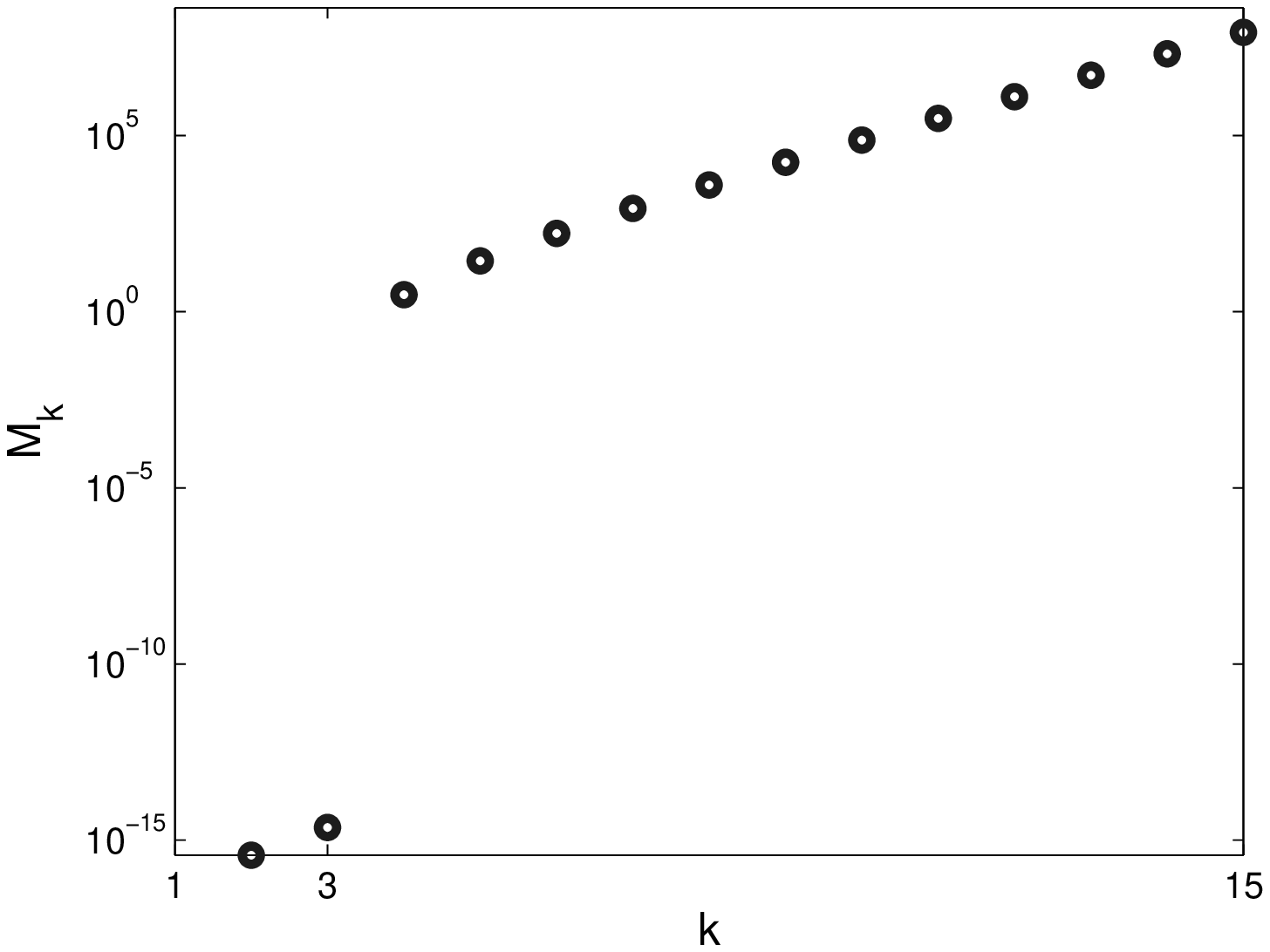,height=4cm}\\[.5cm]
\epsfig{figure=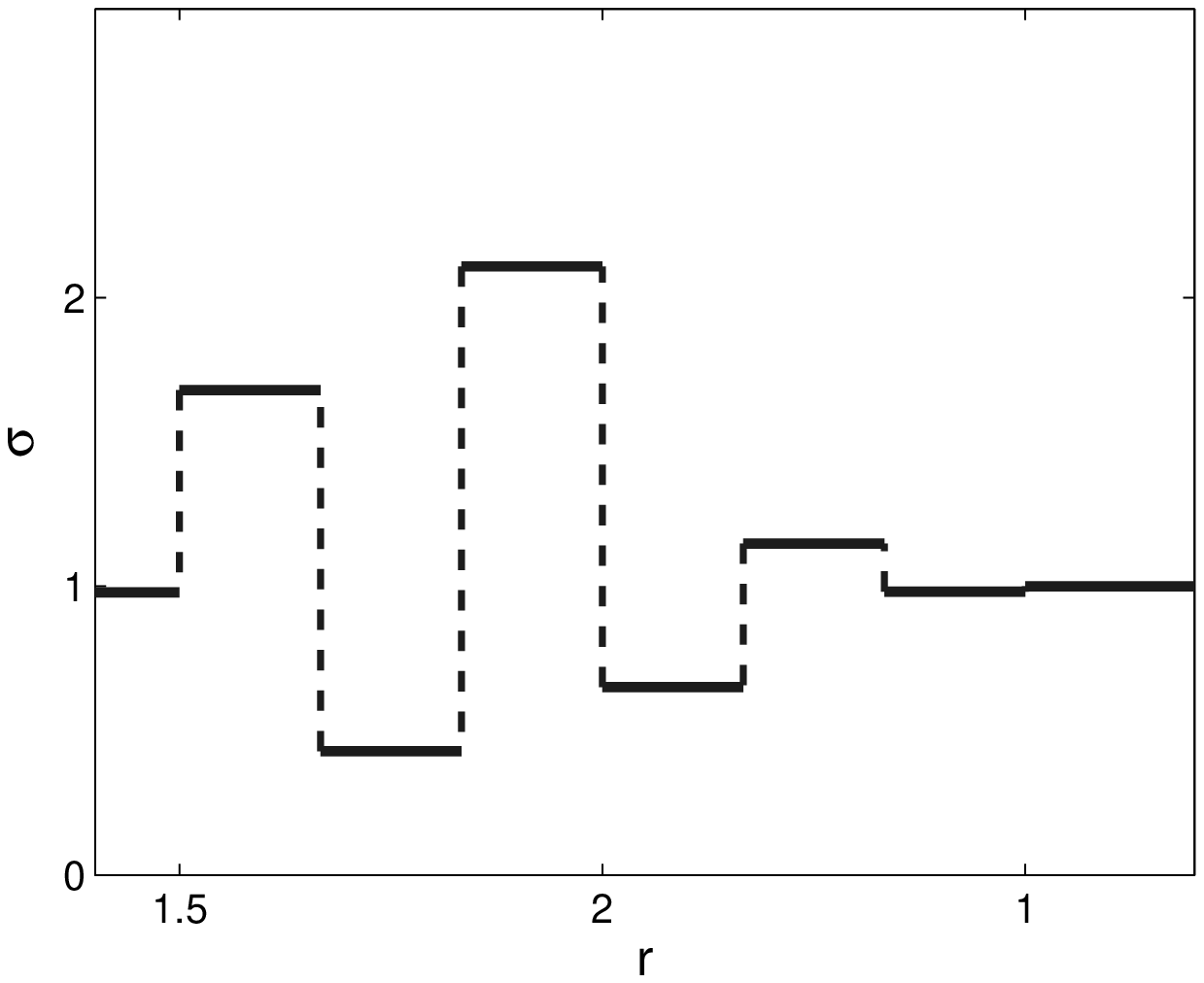, height=4cm}\hskip 1cm
\epsfig{figure=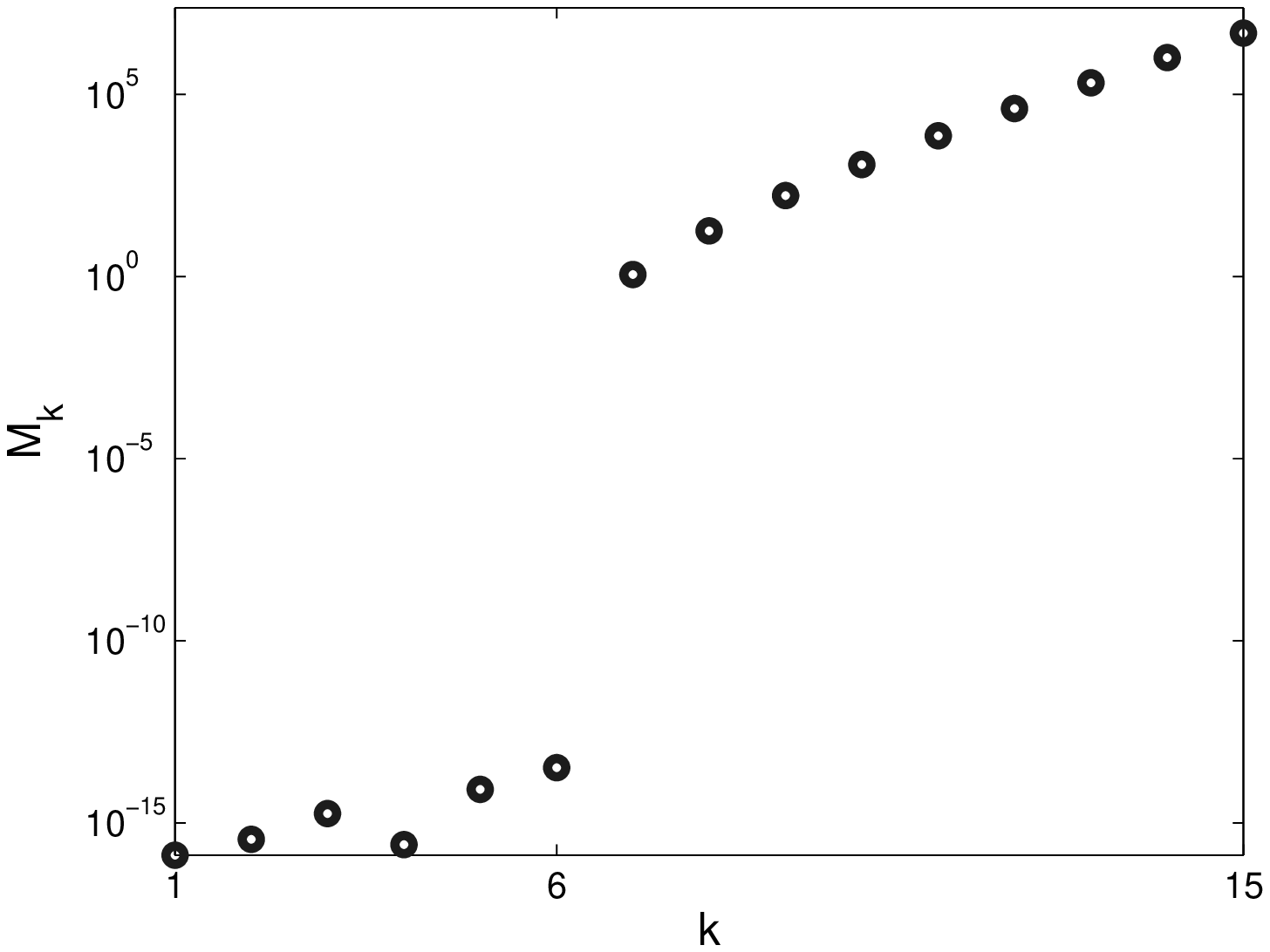,height=4cm}\\[.5cm]
\epsfig{figure=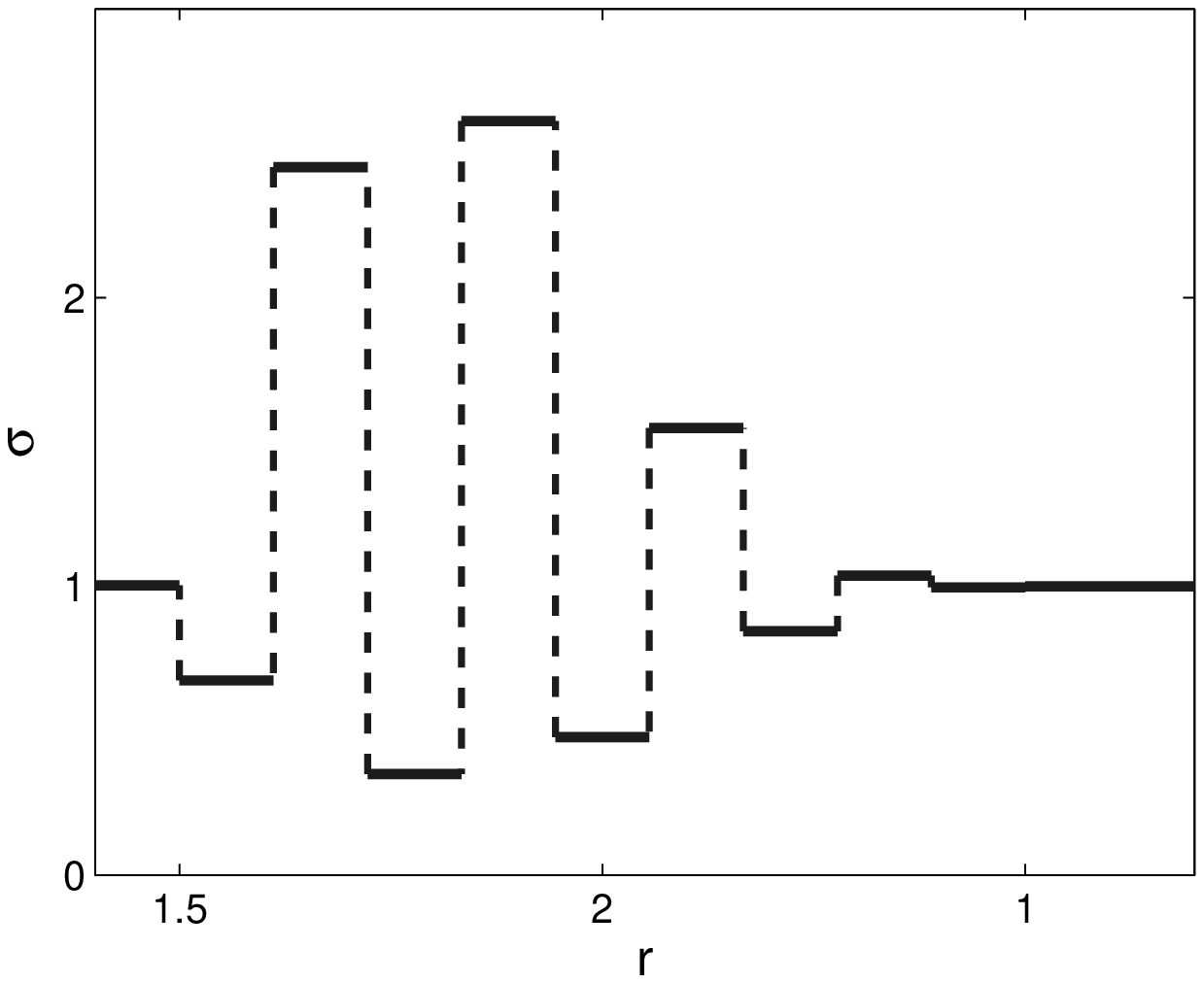, height=4cm}\hskip 1cm
\epsfig{figure=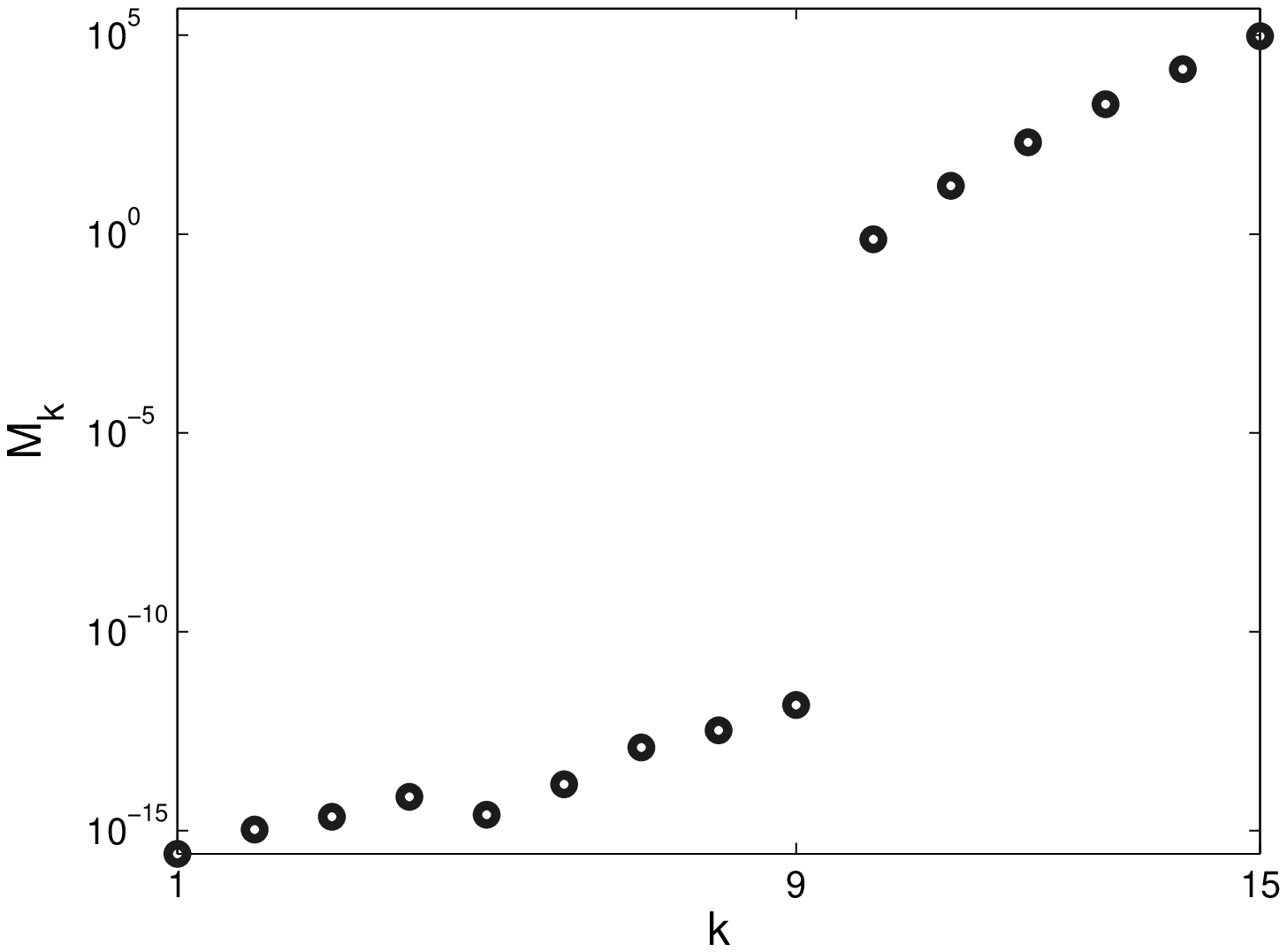,height=4cm}
\end{center}
\caption{Graphs on the left column show the conductivity profile $\Gs$ such that $M_{kk}^{cc}=0$
for $k\leq N$ and plots on the right column show the values of $M_k=M_{kk}^{cc}$ for $N=3, 6,
9$.}\label{fig_cond_nonfix}
\end{figure}

\noindent{\bf Example 1}. Figure \ref{fig_cond_nonfix} shows
computational results of the conductivity $\Gs$ for $N=3,6,9$. It
clearly shows that the larger $N$ is, the more $\Gs$ fluctuates.
One interesting thing to observe is that $\sigma_{N+1}$ takes
values of 1.9695,  0.9791, 1.0029 for $N=3,6,9$, respectively:
they are getting closer to $1$ which is the conductivity of the
exterior part.

\medskip

\begin{figure}[h!]
\begin{center}
\epsfig{figure=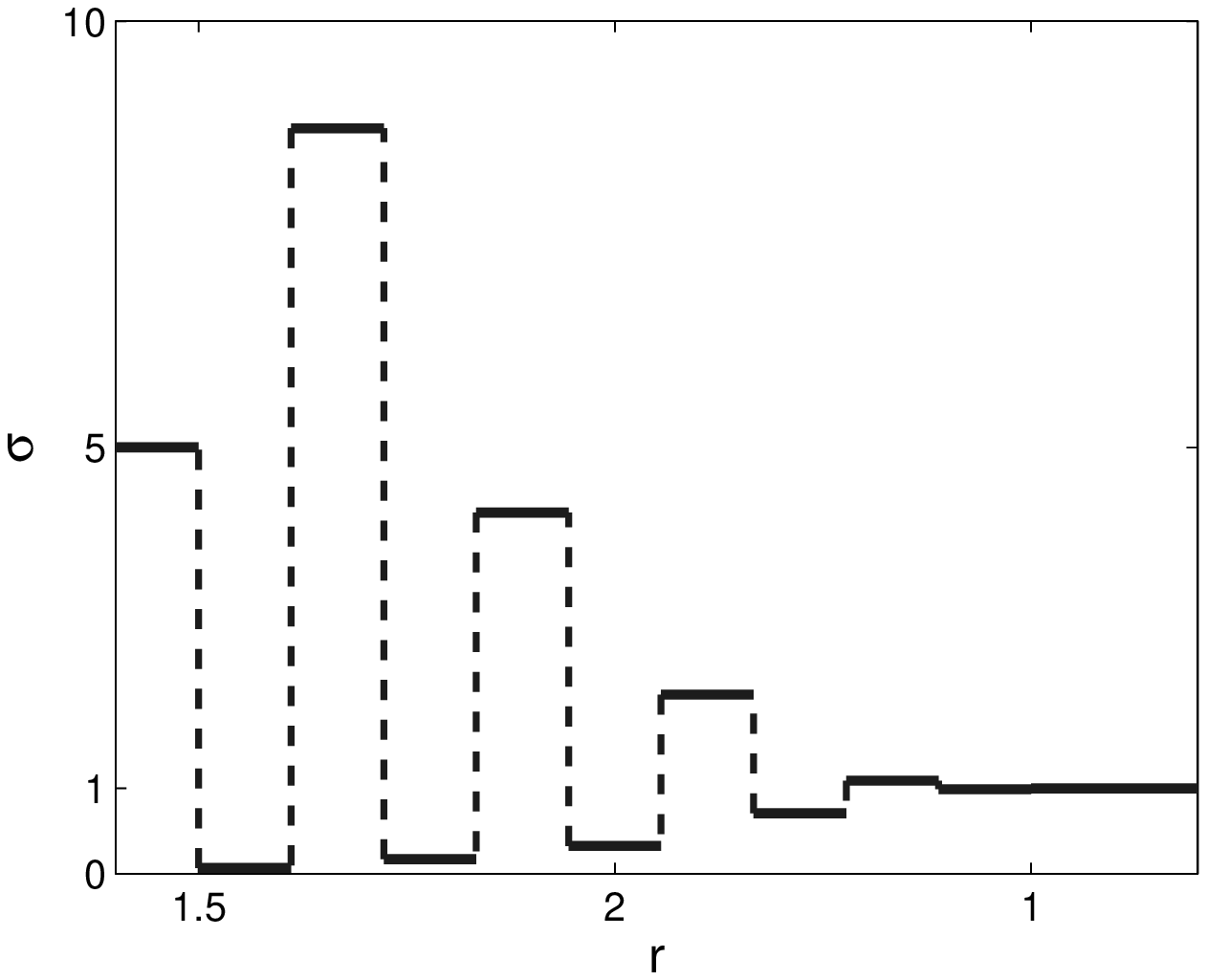, height=4cm}\hskip 1cm
\epsfig{figure=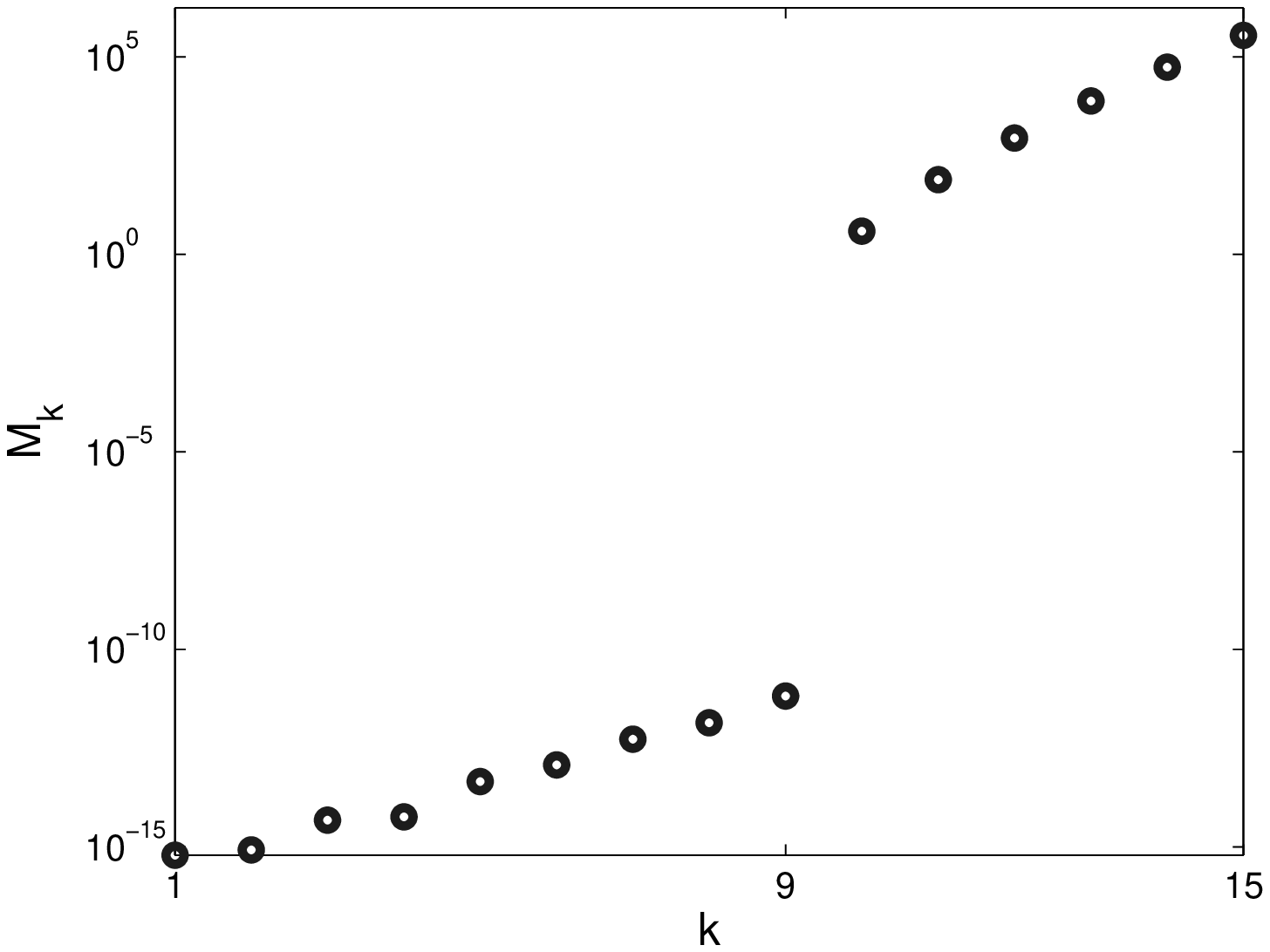,height=4cm} \\[.5cm]
\epsfig{figure=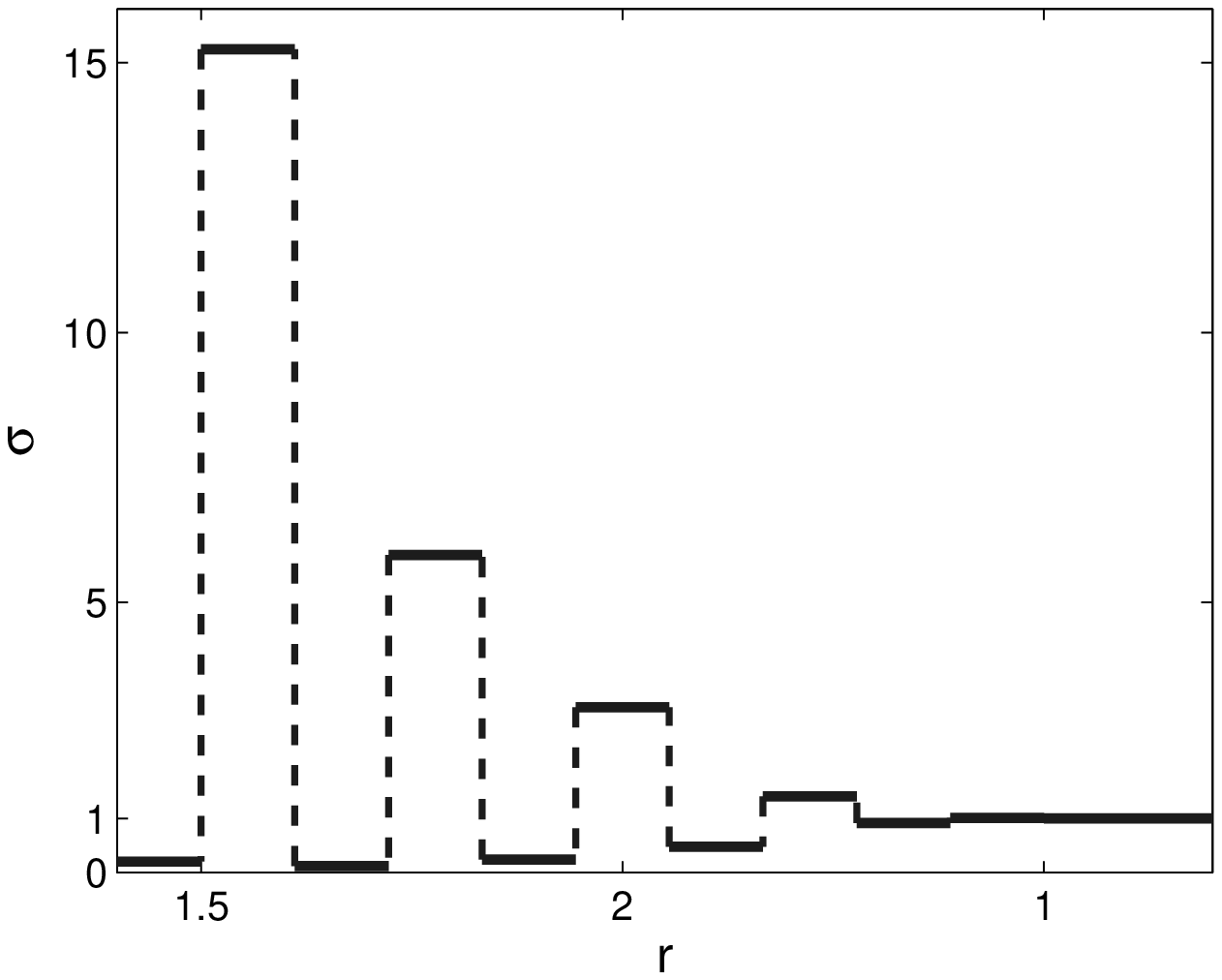, height=4cm}\hskip 1cm
\epsfig{figure=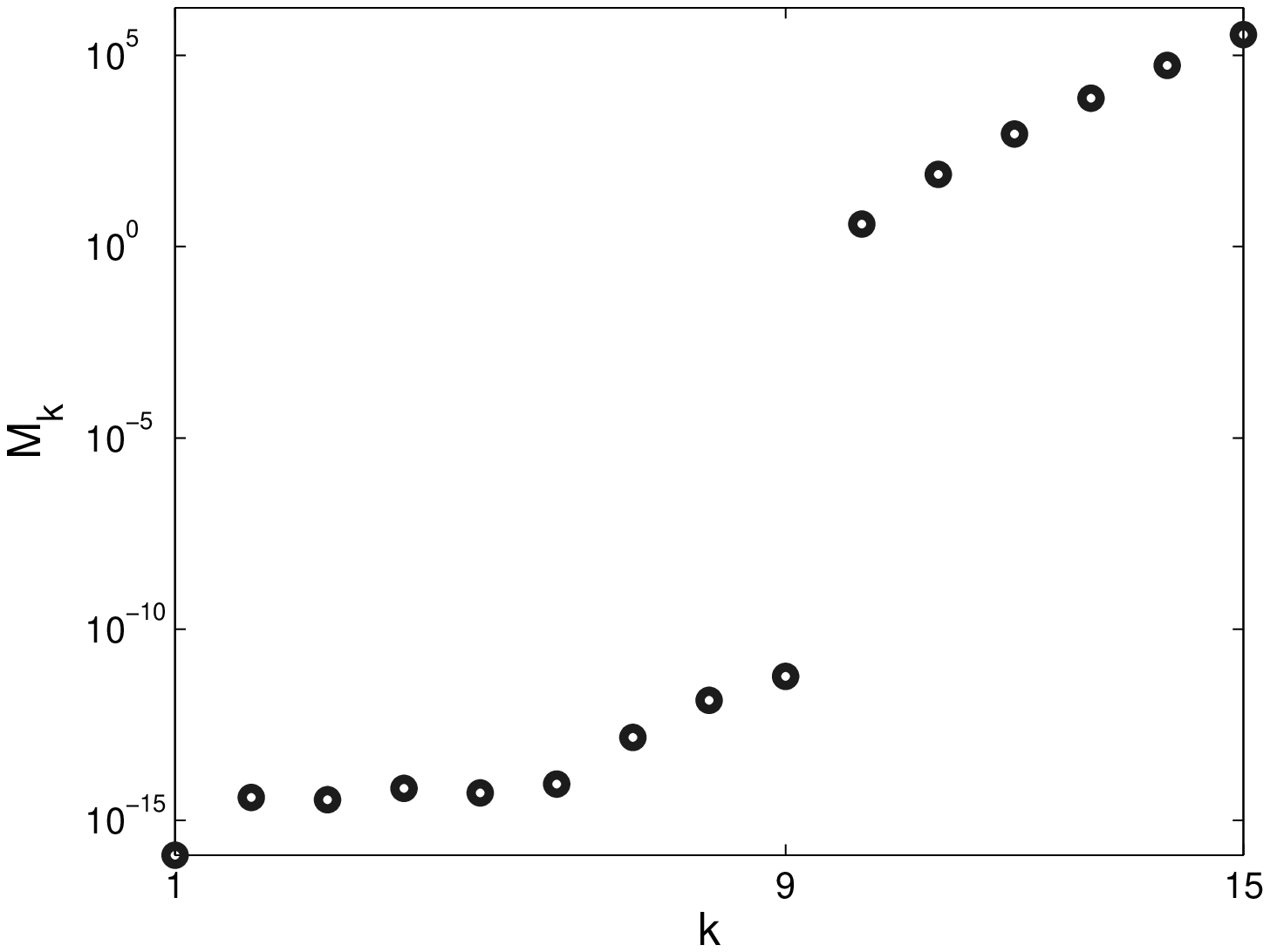,height=4cm}
\end{center}
\caption{Conductivity profile for $N=9$ when $\sigma_{10}$ is fixed. The first row
corresponds to $\sigma_{10}=5$ and the second one to
$\sigma_{10}=0.2$.}\label{fig_corefix}
\end{figure}

\noindent{\bf Example 2}. Here we see what happens if the
conductivity $\sigma_{N+1}$ of the core $D$ is fixed with a value
different from that of exterior, namely 1. We set $N$ (the number
of layers) to be $9$ and $\sigma_{10}=5$ and 0.2. The numerical
results are illustrated in Figure \ref{fig_corefix}. Conductivity
profiles fluctuate more drastically and $M_k$ takes greater values
than those in Example 1.

\medskip

\begin{figure}[h!]
\begin{center}
\epsfig{figure=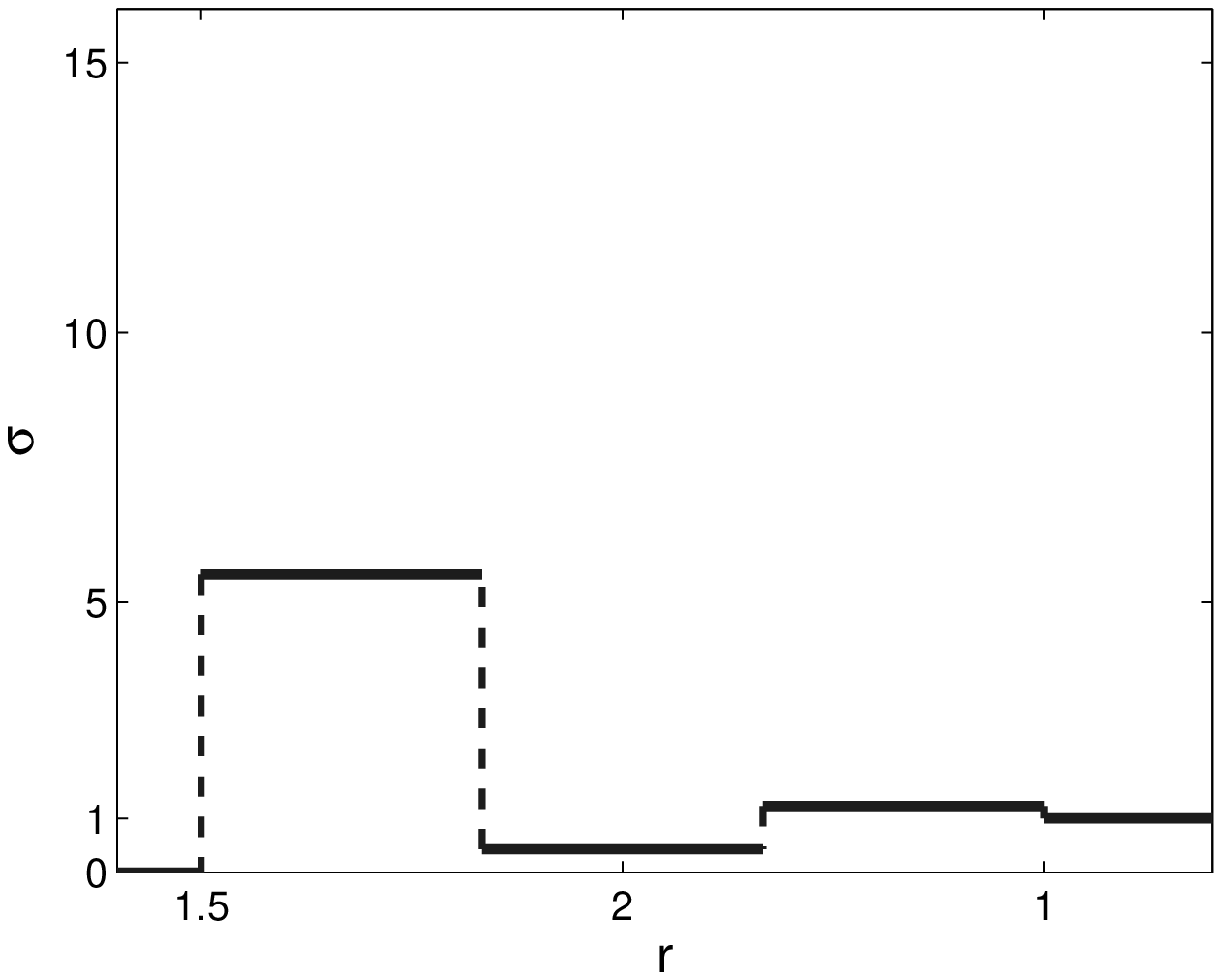, height=4cm}\hskip 1cm
\epsfig{figure=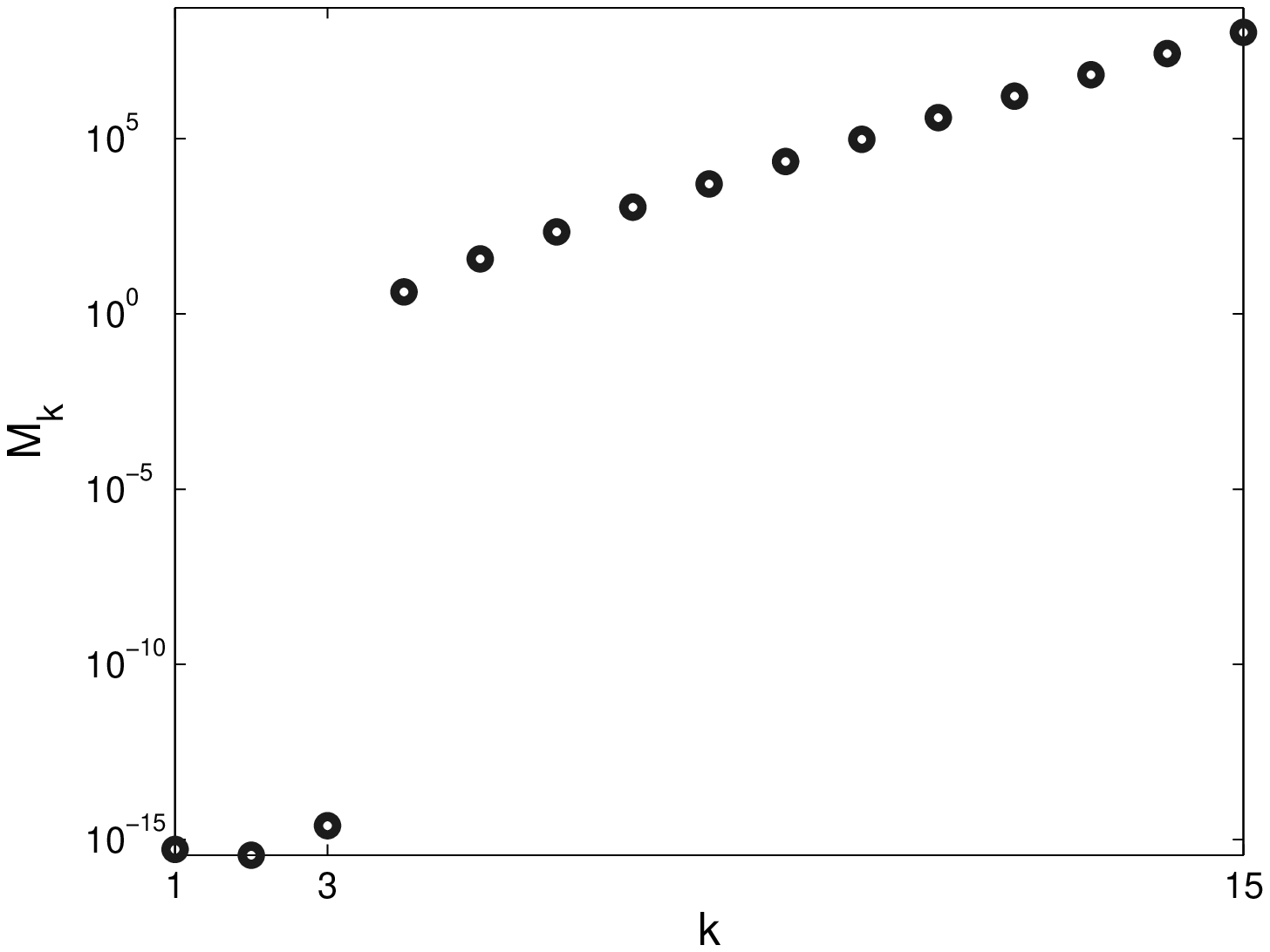,height=4cm}\\[.5cm]
\epsfig{figure=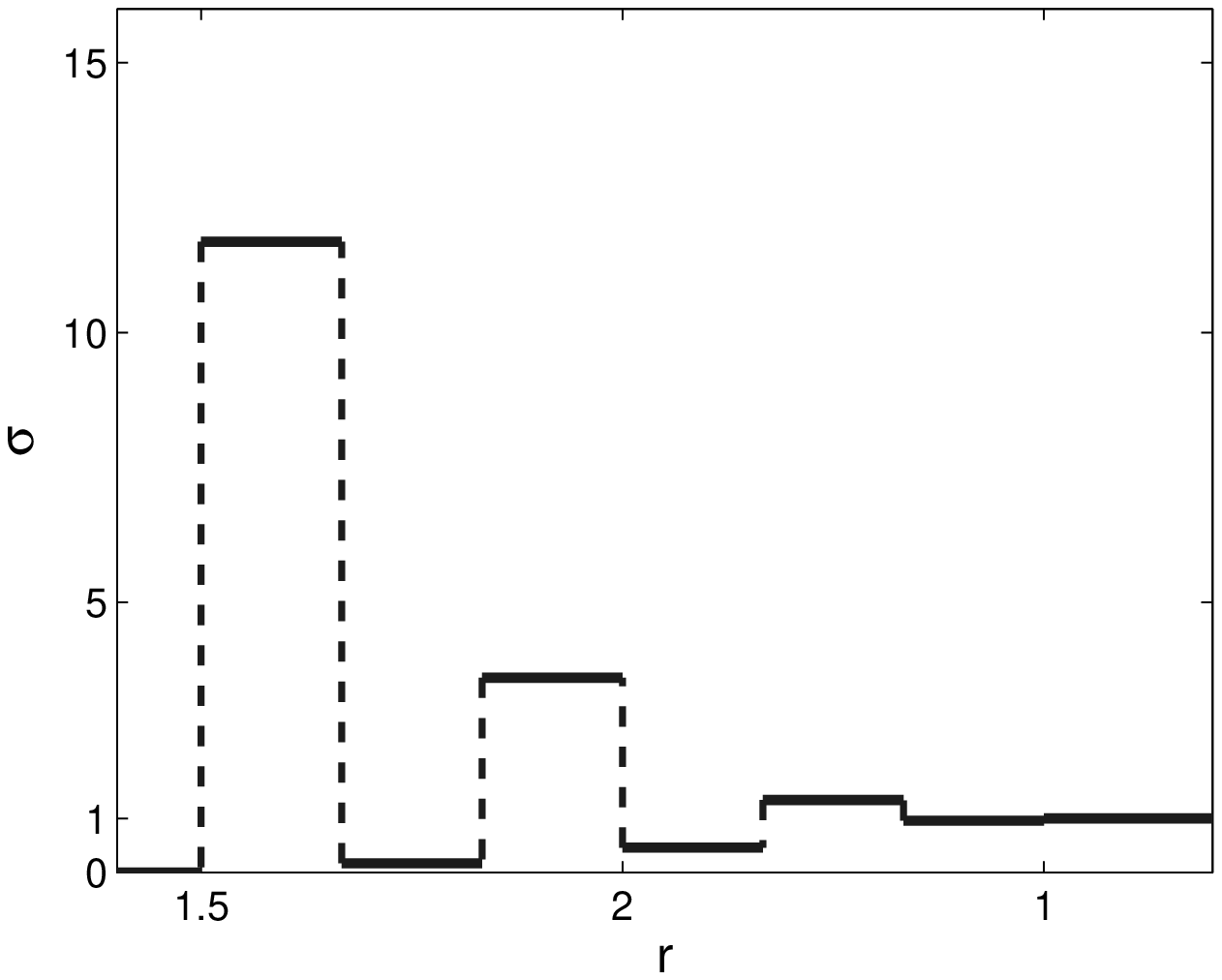, height=4cm}\hskip 1cm
\epsfig{figure=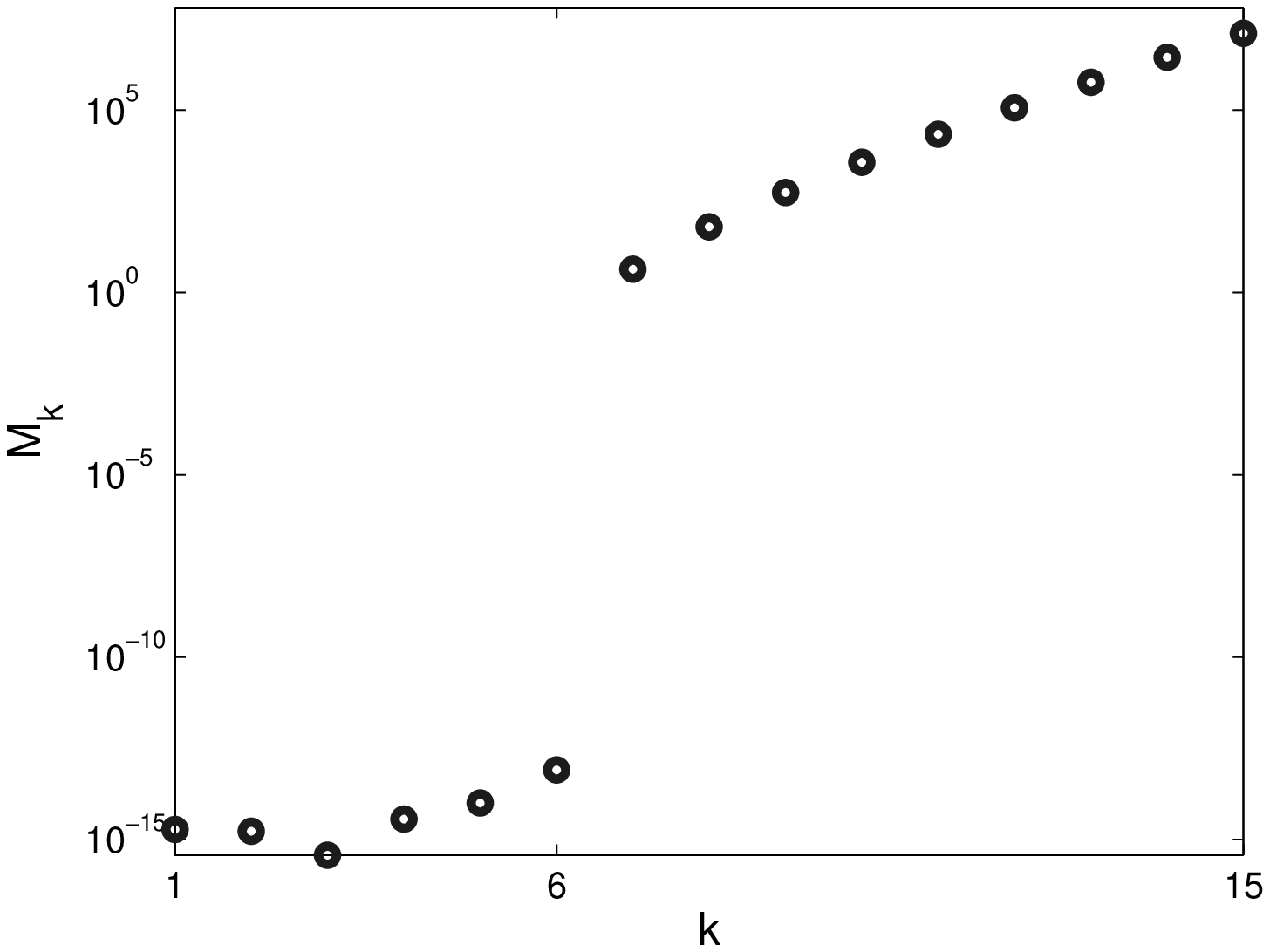,height=4cm}
\end{center}
\caption{Conductivity profile for the near-cloaking with interior conductivity is 0. The first row is when $N=3$
and the second one for $N=6$.}\label{fig_corezero}
\end{figure}

\noindent{\bf Example 3}. This example is for the near-cloaking for which the boundary of the
 core $(r=1)$ is insulated, and hence the conductivity of the core is set to be $0$.
 Figure \ref{fig_corezero} shows the results of computation when $N=3, 6$: the conductivity
 fluctuates on coatings near the core. When $N=3$, the maximal conductivity is 5.5158 and the minimum
 conductivity is 0.4264. When $N=6$, they are
11.6836 and 0.1706.

\section{Conclusion}
We have obtained new near-cloaking examples for the conductivity
problem. We have shown that the GPT-vanishing structures can be
used to enhance the near-cloaking. The GPTs up to the $N$-th order
can be canceled using $(N+1)$ layers with different conductivity
parameters. To make the numerical procedure simple, we have
assumed that the layers are concentric disks centered at the
origin with specific radii. The numerical simulations show that the GPT-vanishing structure exists with these radii.  As mentioned in Introduction, the multi-coating technique can be applied to enhance near-cloaking for
the Helmholtz equation.

\end{document}